\documentclass[onecolumn, floatfix, preprintnumbers, eqsecnum, letterpaper, superscriptaddress, nofootinbib]{revtex4}
\usepackage{graphicx}
\usepackage{microtype}
\usepackage{amsmath}
\usepackage{amssymb}
\usepackage{subfigure}
\usepackage{hyperref}
\usepackage{url}
\usepackage{xcolor}
\usepackage{color}
\usepackage{mathrsfs}
\usepackage{calrsfs}
\usepackage{amsfonts}
\usepackage{eufrak}
\usepackage{tabularx}
\usepackage{eucal}
\usepackage{latexsym}
\usepackage{ragged2e}
\usepackage{epsfig}
\usepackage{textcomp}

\usepackage{caption}
\usepackage{subfigure}

\usepackage[thicklines]{cancel}

\DeclareCaptionJustification{justified}{\leftskip=0pt \rightskip=0pt \parfillskip=0pt plus 1fil}
\definecolor{vividviolet}{rgb}{0.62, 0.0, 1.0}
\definecolor{amaranth}{rgb}{0.9, 0.17, 0.31}
\definecolor{palatinateblue}{rgb}{0.15, 0.23, 0.89}
\definecolor{brightpink}{rgb}{1.0, 0.0, 0.5}
\definecolor{cornflowerblue}{rgb}{0.39, 0.58, 0.93}
\definecolor{deepcarminepink}{rgb}{0.94, 0.19, 0.22}
\definecolor{radicalred}{rgb}{1.0, 0.21, 0.37}

\hypersetup{ linktoc=all,
    colorlinks, linkcolor={palatinateblue},
    citecolor={brightpink}, urlcolor={amaranth}
}

\graphicspath{{Images/}}

\graphicspath{{Images/}}

\def\@fnsymbol#1{\ensuremath{\ifcase#1\or \ddagger \or  $\textleaf$  \or \dagger
\else\@ctrerr\fi}}%

\makeatother

\def\sideremark#1{\ifvmode\leavevmode\fi\vadjust{\vbox to0pt{\vss
 \hbox to 0pt{\hskip\hsize\hskip1em
 \vbox{\hsize1.3cm\tiny\raggedright\pretolerance10000
 \noindent #1\hfill}\hss}\vbox to8pt{\vfil}\vss}}}%

\def\beq{\begin{equation}}
\def\eeq{\end{equation}}

\newcommand{\od}{\mathrm{d}}

\begin{document}

\title{Temperature of McVittie Black Holes in Cavities}

\author{Shi-Bei Kong}
\email{shibeikong@ecut.edu.cn}
\affiliation{School of Science, East China University of Technology, Nanchang 330013, Jiangxi, China}

\author{Yu-Ke Wang}
\email{2024216357@ecut.edu.cn}
\affiliation{School of Nuclear Science and Engineering, East China University of Technology, Nanchang 330013, Jiangxi, China}

\author{Zhen-Bin Zhu}
\email{2022214398@ecut.edu.cn}
\affiliation{School of Science, East China University of Technology, Nanchang 330013, Jiangxi, China}

\begin{abstract}

In this paper, we study the temperature of the McVittie and charged McVittie black hole in cavities. The cavities are isothermal boxes that can make the (charged) McVittie black hole thermodynamically stable. The (charged) McVittie black hole is a dynamical black hole, so the traditional action method can hardly be used to calculate the cavity temperature. As dynamical spacetimes with spherical symmetry, they do not have Killing vectors but have Kodama vectors $K^a$, which are found to be precisely $(\partial/\partial t)^a$, a result that is quite remarkable. We can also obtain their Hawking temperature at the apparent horizons. Based on the Kodama vector and Hawking temperature, we find that the Tolman relation can be used to calculate the cavity temperature if one replaces the Killing vector in the Tolman relation by the Kodama vector for dynamical spacetimes with spherical symmetry. Furthermore, since (charged) de Sitter black hole can be viewed as a special case of the (charged) McVittie black hole, we also obtain the cavity temperature by setting $H^2$ to be $\Lambda/3$, and the results are exactly the same as those from the traditional action method, while our method seems much easier.

\end{abstract}

\maketitle

\section{Introduction}

In previous work \cite{Abdusattar:2022bpg}, we investigated the thermodynamics of the McVittie black hole and found a Hawking-Page like phase transition.
Found in 1933, the McVittie black hole or cosmological black hole \cite{McVittie:1933zz,Faraoni:2007es,Kaloper:2010ec,Nandra:2011ug,Faraoni:2012gz}
can be regarded as a black hole embedded in a spatially flat Friedmann-Robertson-Walker(FRW) universe, which is similar to the de Sitter black hole.
Like the de Sitter black hole, the McVittie black hole has two horizons, the black hole horizon and the cosmological horizon,
both of which can evaporate with different temperature. One can treat the two horizons as two separate thermodynamic systems,
establish the first law of thermodynamics, equation of state, etc. for the two horizons respectively. However, the two kinds of temperature are not equal,
so the McVittie black hole is not in the equilibrium state.

In order to make the McVittie black hole thermodynamically stable, one can place a cavity between the black hole horizon and the cosmological horizon
as is often done for de Sitter black holes \cite{Simovic:2018tdy,Marks:2021fpe}. It has a fixed temperature and allows the black hole to remain in an equilibrium state. In this paper, we use this method to stabilize the McVittie black hole and study its temperature. Unlike the de Sitter black hole,
we cannot integrate the action of the McVittie black hole, because one cannot find a static patch or coordinate system for the McVittie black hole.
Therefore, we cannot obtain the temperature at the horizons from the action. We find that the Hawking temperature and Tolman relation \cite{Rovelli:2010mv} of the McVittie black hole can be used to derive the cavity temperature. The McVittie black hole temperature $T_+$ is defined from the surface gravity at the black hole horizon.
Start from $T_+$, we can obtain the cavity temperature $T_c$ by using the Tolman relation. In a flat spacetime, equilibrium means temperature is the same in every position, but it is not the case for curved spacetime. It is found that in a curved spacetime or a gravitational system, temperature can be different even if equilibrium is established. For a black hole, the temperature of the Hawking radiation cools down as it travels away from
the horizon due to the redshift effect. It has an extreme high temperature near the horizon and is called the Hawking temperature if measured
at an infinite far place. For a stationary black hole, temperature at different distance satisfies the Tolman(or Tolman-Ehrenfest) relation. It says that the multiplication of temperature and the norm of the Killing vector field is a constant, which is actually equivalent to the redshift formula.
However, the McVittie black hole is not a stationary black hole but a dynamical black hole, so it does not have a Killing vector field. Fortunately,
for a dynamical black hole, it has a Kodama vector $K^a$ as a counterpart of the Killing vector. We obtain the Kodama vector for the McVittie black hole
and find that it is just $(\partial/\partial t)^a$, which is very remarkable and interesting. In the following, we put the Hawking temperature and Kodama vector into the Tolman relation and obtain the cavity temperature at $r_c$. What's more, if the Hubble parameter is treated as a constant, i.e. $H^2=\Lambda/3$, the McVittie black hole becomes the de Sitter black hole. Therefore, from our result we also obtain the cavity temperature of the de Sitter black hole. Interestingly, the result is the same as that from the traditional action method. We further investigate the charged McVittie black hole \cite{Gao:2004cr,Faraoni:2014nba,Guariento:2019ock} and find that the Kodama vector is still $(\partial/\partial t)^a$.
We also obtain the cavity temperature of the charged McVittie black hole from the Tolman relation and Kodama vector. As a special case of the charged McVittie black hole, i.e. charged de Sitter black hole, we also obtain its cavity temperature by setting $H^2=\Lambda/3$, and the result is also the same as that from the action method. These results could be regarded as a manifestation of consistency, but our method is much easier.
 
The paper is organized in the following way. In Sec.II, we introduce the McVittie black hole and calculate its temperature 
at the cavity. In Sec.III, we introduce the charged McVittie black hole and calculate its temperature 
at the cavity. In Sec.IV, we give the conclusions and discussions.

\section{Cavity Temperature of the McVittie Black Hole}

The metric of the McVittie black hole can be written as \cite{Nandra:2011ug}

\begin{alignat}{1}
\od s^2=-\left(1-\frac{2m}{r}-H^2r^2\right)\od t^2-\frac{2Hr}{\sqrt{1-\frac{2m}{r}}}\od t\od r+\frac{1}{1-\frac{2m}{r}}\od r^2
+r^2(\od\theta^2+\sin^2\theta\od\varphi^2), \label{metric}
\end{alignat}
where $H\equiv\dot{a}(t)/a(t)$ is the Hubble parameter\footnote{We use different signature with the original paper \cite{Nandra:2011ug}.}.
For convenience, we use $h_{ab}$ to denote the metric of the $(t,r)$ part and also give its inverse metric
\begin{alignat}{1}
(h^{ab})=\left(
               \begin{array}{cc}
                 -\frac{1}{1-\frac{2m}{r}}, & -\frac{Hr}{\sqrt{1-\frac{2m}{r}}} \\ \\ 
                 -\frac{Hr}{\sqrt{1-\frac{2m}{r}}}, &1-\frac{2m}{r}-H^2r^2       
               \end{array}
        \right),
\end{alignat}
where $a,b=0,1$ and $x^0=t,x^1=r$.

By definition, the apparent horizon of a dynamical spacetime with spherical symmetry satisfies the following condition \cite{Hayward:1994bu,Cai:2008gw}
\begin{alignat}{1}
h^{ab}\nabla_ar\nabla_br=\nabla^ar\nabla_ar=0, \label{ahd}
\end{alignat}
so for the McVittie black hole (\ref{metric}), this condition can be written as
\begin{alignat}{1}
1-\frac{2m}{r}-H^2r^2=0. \label{ah}
\end{alignat}
It has two solutions generally, and we denote the smaller one as $r_+$, i.e. the black hole horizon, and the larger one as 
$r_{cosmo}$, i.e. the cosmological horizon.
From the above condition, one can obtain a useful relation 
\begin{alignat}{1}
\dot{r}_+=\frac{2 H \dot{H} r_{+}^{3}}{1-3 H^2 r_{+}^2}. \label{dotr}
\end{alignat}

For a dynamical spacetime with spherical symmetry, its Killing vector is missing, but it has a Kodama vector \cite{Cai:2008gw} that plays a similar role.
The Kodama vector is defined and obtained as follows
\begin{alignat}{1}
K^a\equiv&-\epsilon^{ab}\nabla_b r=\delta_{0}^a=\left(\frac{\partial}{\partial t}\right)^a, 
\end{alignat} 
which is exactly  $(\partial/\partial t)^a$ and very remarkable, see Appendix for the calculation.
We have $K_aK^a=h_{ab}K^aK^b=h_{00}$, so it is timelike in $r_+<r<r_{cosmo}$, null in $r=r_+$ or $r_{cosmo}$, and spacelike in $r<r_+$ 
or $r>r_{cosmo}$, respectively. In the timelike region, its norm is 
\begin{alignat}{1}
||K^a||=\sqrt{|K_aK^a|}=\sqrt{-h_{00}}=\sqrt{1-\frac{2m}{r}-H^2r^2}. \label{norm}
\end{alignat}
Kodama vector plays an important role in the establishment of thermodynamics for dynamical black holes and the FRW universe.
By using it, Hayward proposed a unified first law \cite{Hayward:1997jp}, which results to the first law of thermodynamics after projection
on the apparent horizon.

Corresponding to the Kodama vector, one can define surface gravity at radius $r$ for a dynamical spacetime with 
spherical symmetry \cite{Hayward:1997jp,Cai:2006rs}
\begin{alignat}{1}
\kappa:=\frac{1}{2}\nabla^a\nabla_ar=\frac{1}{2\sqrt{-h}}\frac{\partial}{\partial x^{a}}\left(\sqrt{-h}~h^{ab}\frac{\partial r}{\partial x^{b}}\right), \label{sg}
\end{alignat}
where $h$ is the determinant of $h_{ab}$, and inserting the metric (\ref{metric}), one can obtain 
\begin{alignat}{1}
\kappa=\frac{m}{r^2}-\frac{\dot{H}r}{2\sqrt{1-\frac{2m}{r}}}-H^2r.
\end{alignat}
At the black hole horizon $r_+$, the surface gravity can be written as 
\begin{alignat}{1}
\kappa_+=\frac{1}{2r_{+}}\left(1-3H^2r_{+}^2\right)\left(1-\frac{\dot{r}_{+}}{2 H^2r_+^2}\right),
\end{alignat}
where (\ref{ah}) is used, so one can obtain the Hawking temperature of the black hole horizon 
\begin{alignat}{1}
T_+=\frac{1}{4\pi r_+}(1-3H^2r_+^2)\left(1-\frac{\dot{r}_{+}}{2H^2r_+^2}\right).
\end{alignat}

At the cavity with radius $r_+<r_c<r_{cosmo}$, the temperature $T_c$ is fixed with the temperature of the Hawking radiation at that place,
which is different from $T_+$ but can be obtained from the Tolman relation. Tolman relation \cite{Rovelli:2010mv} says that at a finite place of a stationary black hole, its temperature has the following relation with the Hawking temperature
\begin{alignat}{1}
T=\frac{T_H}{||\xi||},
\end{alignat}
where $\xi$ is the Killing vector of the black hole. However, the McVittie black hole is not a stationary black hole and does not have a Killing vector, 
so this relation cannot be directly used to calculate the temperature at the cavity. Fortunately, we have a counterpart for the Killing vector, 
i.e. the Kodama vector, which plays a similar role for dynamical black holes like the McVittie black hole. We can put the norm (\ref{norm}) of the Kodama vector for the McVittie black hole in the Tolman relation and obtain the cavity temperature at the cavity $r=r_c$ 
\begin{alignat}{1}
T_c=\frac{T_+}{||K^a||_c}=\frac{(1-3H^2r_+^2)\left(1-\frac{\dot{r}_{+}}{2H^2r_+^2}\right)}
{4\pi r_{+}\sqrt{1-\frac{2m}{r_c}-H^2r_c^2}}
=\frac{(1-3H^2r_+^2)\left(1-\frac{\dot{r}_{+}}{2H^2r_+^2}\right)}
{4 \pi r_{+}\sqrt{\left(1-\frac{r_+}{r_c}\right)[1-H^2(r_+^2+r_+r_c+r_c^2)]}}.
\end{alignat}
The McVittie black hole (\ref{metric}) can reduce to de Sitter black hole by fixing $H^2=\Lambda/3$,
and in this case we have $\dot{r}_+=0$, see (\ref{dotr}). Therefore, from the above result one can also obtain the cavity temperature of the de Sitter black hole,
\begin{alignat}{1}
T_c=\frac{1-\Lambda r_+^2}{4 \pi r_{+}\sqrt{\left(1-\frac{r_+}{r_c}\right)\left[1-\frac{\Lambda}{3}(r_+^2+r_+r_c+r_c^2)\right]}},
\end{alignat}
which is exactly the same as the result from the action method \cite{Simovic:2018tdy}. This could be regarded as a manifestation of consistency and
shows that the above method is reasonable.

\section{Cavity Temperature of the Charged McVittie Black Hole}

Similar to the charged de Sitter black hole, the solution of the charged McVittie has also been found.
The metric of the charged McVittie black hole can be written as \cite{Guariento:2019ock}
\begin{alignat}{1}
\od s^2=-\left(N^2-H^2r^2\right)\od t^2-\frac{2Hr}{N}\od t\od r+\frac{1}{N^2}\od r^2+r^2(\od\theta^2+\sin^2\theta\od\varphi^2), \label{metric2}
\end{alignat}
where $N\equiv\sqrt{1-2m/r+Q^2/r^2}, H\equiv\dot{a}(t)/a(t)$, and $Q$ is the charge.

The apparent horizon of the charged McVittie black hole also satisfies (\ref{ahd}), which in this case can be written as
\begin{alignat}{1}
1-\frac{2m}{r}+\frac{Q^2}{r^2}-H^2r^2=0, 
\end{alignat}
and the smaller solution is the black hole horizon denoted by $r_+$.
From the above condition, we can also obtain
\begin{alignat}{1}
\dot{H}=-(r_+^2-3mr_++2Q^2)\frac{\dot{r}_+}{Hr_+^5}.  \label{doth}
\end{alignat}

It can be proved that the Kodama vector for the charged McVittie spacetime is still $K^a=(\partial/\partial t)^{a}$,
which is very shocking, see Appendix for the proof. In the timelike region, its norm is 
\begin{alignat}{1}
||K^a||=\sqrt{|K_aK^a|}=\sqrt{-h_{00}}=\sqrt{1-\frac{2m}{r}+\frac{Q^2}{r^2}-H^2r^2}.
\end{alignat}
Insert the metric (\ref{metric2}) into the definition (\ref{sg}) of the surface gravity, 
one obtains the surface gravity at radius $r$ corresponding to the Kodama vector of the charged McVittie black hole,
\begin{alignat}{1}
\kappa=\frac{m}{r^2}-\frac{Q^2}{r^3}-\frac{\dot{H}r}{2N}-H^2r.
\end{alignat}
By using (\ref{ah}), the surface gravity at the black hole horizon can be written as
\begin{alignat}{1}
\kappa_+=\frac{1}{2r_+}\left(1-3H^2r_+^2-\frac{Q^2}{r_+^2}\right)\left(1-\frac{\dot{r}_+}{2H^2r_+^2}\right),
\end{alignat}
so the Hawking temperature can be obtained 
\begin{alignat}{1}
T=\frac{\kappa_+}{2\pi}=\frac{1}{4\pi r_+}\left(1-3H^2r_+^2-\frac{Q^2}{r_+^2}\right)\left(1-\frac{\dot{r}_+}{2H^2r_+^2}\right).
\end{alignat}

Therefore, using the Tolman relation, the temperature at the cavity radius $r_c$ for the charged McVittie black hole is given by
\begin{alignat}{1}
T_c=&\frac{T_+}{||K^a||_c}=\frac{T}{\sqrt{-h_{00}|_c}}=\frac{\left(1-3H^2r_+^2-\frac{Q^2}{r_+^2}\right)\left(1-\frac{\dot{r}_+}{2H^2r_+^2}\right)}
{4\pi r_+\sqrt{1-\frac{2m}{r_c}+\frac{Q^2}{r_c^2}-H^2r_c^2}}
\nonumber \\
=&\frac{\left(1-3H^2r_+^2-\frac{Q^2}{r_+^2}\right)\left(1-\frac{\dot{r}_+}{2H^2r_+^2}\right)}
{4\pi r_+\sqrt{(1-\frac{r_+}{r_c})\left[1-\frac{Q^2}{r_+r_c}-H^2(r_c^2+r_+^2+r_c r_+)\right]}}.
\end{alignat}
If one sets $H^2=\Lambda/3$, the charged McVittie black hole reduces to the charged de Sitter black hole, 
so one can obtain the cavity temperature of the charged de Sitter black hole
\begin{alignat}{1}
T_c=\frac{1-\Lambda r_+^2-\frac{Q^2}{r_+^2}}
{4\pi r_+\sqrt{(1-\frac{r_+}{r_c})\left[1-\frac{Q^2}{r_+r_c}-\frac{\Lambda}{3}(r_c^2+r_+^2+r_c r_+)\right]}}, \label{tcdsc}
\end{alignat}
which is still the result from the action method \cite{Simovic:2018tdy}.
Again, it shows that the above method is consistent and reasonable.

\section{Conclusions and Discussions}

In this paper, we obtain the temperature of the McVittie black hole and the charged McVittie black hole in a cavity with radius $r_c$.
The cavity temperature is derived from the black hole Hawking temperature and Kodama vector by using the Tolman relation. 
We also obtain the cavity temperature of the de Sitter black hole and the charged de Sitter black hole respectively by setting $H^2=\Lambda/3$
from the above results, which are the same as those from the action method. By the way, we also find that the Kodama vector of the McVittie and charged
McVittie black hole is just $\partial/\partial t$, which is very remarkable and interesting.

It shows that this method to obtain the cavity temperature is consistent and applicable. It is obvious that this method is much easier than the action method and can be used to similar investigations. It also shows that Kodama vector is very important and useful for dynamical spacetimes.

\section*{Acknowledgments}

This work is supported by National Natural Science Foundation of China (NSFC) under grant 12465011
and East China University of Technology (ECUT) under grant DHBK2023002.

\appendix

\section{Kodama Vector}

The Kodama vector for a dynamical spacetime with spherical symmetry is defined as
\begin{alignat}{1}
K^a:=-\epsilon^{ab}\nabla_b r,
\end{alignat} 
and for the McVittie and charged McVittie cases, it is just $(\partial/\partial t)^a$.
Here is the proof:
\begin{alignat}{1}
K^a=&-\epsilon^{ab}\nabla_b r=-h^{ac}h^{bd}\epsilon_{cd}\delta_{b1}=-h^{ac}h^{1d}\epsilon_{cd}
=-h^{0c}h^{1d}\epsilon_{cd}\delta_{0}^a
\nonumber \\
=&-(h^{00}h^{11}\epsilon_{01}+h^{01}h^{10}\epsilon_{10})\delta_{0}^a
=-(h^{00}h^{11}-h^{01}h^{10})\epsilon_{01}\delta_{0}^a
\nonumber \\
=&-\det{h^{ab}}\sqrt{|\det{h_{ab}}|}\delta_{0}^a
=-(\det{h_{ab}})^{-1}\sqrt{-\det{h_{ab}}}\delta_{0}^a
\nonumber \\
=&\frac{\delta_{0}^a}{\sqrt{-\det{h_{ab}}}}=\delta_{0}^a=\left(\frac{\partial}{\partial t}\right)^a,
\end{alignat} 
where $\epsilon_{cd}=\sqrt{|h_{ab}|}(\od t)_c\wedge(\od r)_d$.

\end{document}